\documentclass[aps,prl,twocolumn]{revtex4}

\usepackage{graphicx}
\usepackage{dcolumn}
\usepackage{bm}
\usepackage{bezier}

\usepackage{url}

\usepackage{color}

\usepackage{amsmath}

\DeclareMathOperator{\Imag}{Im}

\begin{document}

\title{Topology-controlled thermopower oscillations in multiterminal Andreev interferometers}
\author{Pavel E. Dolgirev}
\affiliation{Skolkovo Institute of Science and Technology, Skolkovo Innovation Center, 3 Nobel St., 143026 Moscow, Russia}
\author{Mikhail S. Kalenkov}
\affiliation{I.E. Tamm Department of Theoretical Physics, P.N. Lebedev Physical Institute, 119991 Moscow, Russia}
\affiliation{Moscow Institute of Physics and Technology, Dolgoprudny, 141700 Moscow region, Russia}

\author{Andrei D. Zaikin}
\email{andrei.zaikin@kit.edu}
\affiliation{Institut f{\"u}r Nanotechnologie, Karlsruher Institut f{\"u}r Technologie (KIT), 76021 Karlsruhe, Germany}
\affiliation{National Research University Higher School of Economics, 101000 Moscow, Russia}

\date{\today}

\begin{abstract}
We theoretically investigate coherent oscillations of the thermopower $\mathcal{S}$ as a function of the magnetic flux $\Phi$ in six-terminal Andreev interferometers. We demonstrate that the thermopower behavior is determined
by a number of contributions originating from the Josephson-like and Aharonov-Bohm-like effects as well as from electron-hole asymmetry. The relative weight of these contributions depends on the relation between temperature, voltage bias and an effective Thouless energy of our setup. We particularly emphasize the role of the system topology that may have a dramatic impact on the behavior of $\mathcal{S}(\Phi)$.
\end{abstract}

\maketitle

\section{Introduction}

Long-range quantum coherence of electrons in normal-superconducting (NS) hybrid nanostructures gives rise to a large number of intriguing phenomena \cite{belzig1999quasiclassical} which can be directly observed at sufficiently low temperatures. A list of such phenomena -- by far incomplete -- includes the so-called $\pi$-junction state in SNS structures controlled by driving electrons in the N-metal out of equilibrium \cite{V,WSZ,Yip,Teun}, conductance reentrance  \cite{SN96,GWZ97}, proximity-induced Aharonov-Bohm effect in mesoscopic SN-rings \cite{Petrashov_add,SN96,GWZ97,nakano1991quasiparticle,Grenoble}, non-local Andreev reflection \cite{BF,BeckmannCAR,TeunCAR,VenkatCAR,GZCAR,KZCAR,GKZ} and strongly enhanced thermoelectric effect due to spin-dependent electron scattering \cite{Kalenkov12,Machon,Ozaeta,KZ14,KZ15,Beckmann} or Andreev reflection at different NS-interfaces \cite{V2,VH,KZ17}. The latter mechanism could be responsible for large thermoelectric signal observed in Andreev interferometers of different topology \cite{Venkat1,Venkat2,Petrashov03,Venkat3,Petrashov16}.

Note that the thermopower detected in such interferometers exhibits coherent oscillations as a function of
the external magnetic flux $\Phi$ with the period equal to the flux quantum $\Phi_0=\pi /e$ \cite{C1}. Depending on the sample topology these oscillations can be described by either odd or even functions of $\Phi$  \cite{Venkat1}. Increasing an external bias one could also observe these oscillations to vanish and then re-appear with the phase shift equal to $\pi$ \cite{Petrashov03}. In our recent work \cite{PD18} we offered an elaborate explanation for these striking observations. In particular, we argued that the low temperature behavior of Andreev interferometers is essentially determined by a trade-off between non-equilibrium Josephson and Aharonov-Bohm effects resulting in phase-coherent periodic oscillations of both electric currents flowing between different terminals and the thermopower as functions of the magnetic flux $\Phi$. While in general these oscillations were found to be neither even nor odd in $\Phi$, they can reduce to both these particular limits depending on the relation between external voltage bias $eV$ and/or temperature $T$ and the characteristic Thouless energy of the device ${\cal E}_{\rm Th}$ to be defined below. The physics of the $\pi$-shifted thermopower oscillations observed in experiments \cite{Petrashov03} was found to be similar to that elucidated before \cite{V,WSZ,Yip,Teun} for the Josephson $\pi$-junction state controlled by an external voltage bias $V$.

In this Letter we will further extend our analysis by considering Andreev interferometers of a different topology as compared to that addressed in 
\cite{PD18}, cf. Fig.~\ref{fig: geom}. We will explicitly evaluate the flux- or phase-dependent thermopower and demonstrate that the system topology 
-- in addition to such parameters as $eV$, $T$ and ${\cal E}_{\rm Th}$ -- essentially determines its thermoelectric properties in the quantum limit.

\section{The model and basic formalism}
\label{sec: Formalism}

\begin{figure}[!ht]
\includegraphics[width=0.7\columnwidth]{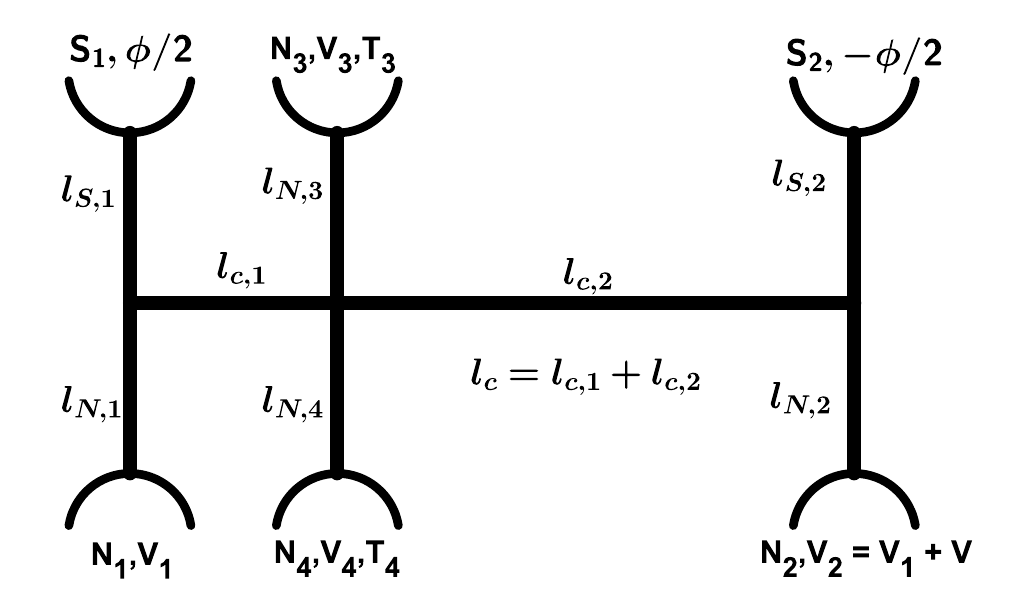}
\caption{Schematics of a six-terminal Andreev interferometer consisting of two superconducting and four normal terminals interconnected by normal metallic wires of different lengths.}
\label{fig: geom}
\end{figure}

In what follows we will consider a multiterminal hybrid structure schematically depicted in Fig.~\ref{fig: geom}.
This structure consists of two superconducting and four normal terminals interconnected by normal diffusive wires of equal cross section ${\cal A}$ and different lengths as shown in Fig.~\ref{fig: geom}. The superconducting terminals S$_1$ and S$_2$ are described by the complex order parameter $\Delta e^{\pm i\phi/2}$, i.e. the phase difference between them is set equal to $\phi$. In practice, this phase difference can be controlled, e.g., in a superconducting loop geometry encircling the magnetic flux $\Phi$, in which case one has $\phi = 2\pi \Phi/\Phi_0$. Two normal terminals N$_1$ and N$_2$ are attached to the voltage source $V=V_2-V_1$, while two remaining terminals N$_3$ and N$_4$ are electrically isolated from any external circuit and are kept at different temperatures $T_3$ and $T_4$. Accordingly, the thermoelectric potentials $V_3$ and $V_4$ are generated at these two normal terminals.

An important energy scale of our device is controlled by the Thouless energy ${\cal E}_{\rm Th} = D/L^2$, where $D$ is the wire diffusion coefficient and $L = l_{S,1} + l_c + l_{S,2}$ is effective distance between the two superconducting terminals. Provided this distance strongly exceeds the superconducting coherence length we have ${\cal E}_{\rm Th} \ll \Delta$, i.e. the relevant energy scale in our problem stays well below the superconducting gap.

In order to proceed we will make use of the standard quasiclassical formalism of the superconductivity theory~\cite{belzig1999quasiclassical}. Here we will adopt the standard $\theta$-parameterisation of the quasiclassical Green functions \cite{ZZh,belzig1999quasiclassical}
\begin{equation}
\hat{G}^R = \begin{pmatrix}
{\cal G}_{11} & {\cal F}_{12}\\
{\cal F}_{21} & {\cal G}_{22}
\end{pmatrix} = \begin{pmatrix}
\cosh \theta & e^{i\chi} \sinh \theta \\
-e^{-i\chi} \sinh\theta  & -\cosh\theta
\end{pmatrix},
\end{equation}
where $\hat{G}^R$ is $2\times 2$ matrix in the Nambu space representing the retarded quasiclassical Green function, $\theta$ and $\chi$ are complex functions satisfying the following spectral Usadel equations
\begin{eqnarray}
D \Delta \theta & = & -2 i \epsilon \sinh \theta + \frac{1}{2} D (\nabla \chi)^2 \sinh 2\theta \label{eq: Theta}\\
\nabla j_E & = & 0,\ j_E = \sinh^2 \theta \cdot \nabla \chi.
\label{eq: chi}
\end{eqnarray}

The kinetic equations can be cast to the following standard form:
\begin{align}
& \nabla j_L = 0 ,\ j_L =  D_L \nabla f_L - \mathcal{Y} \nabla f_T + j_s f_T \label{eq: j_L}\\
& \nabla j_T = 0 ,\ j_T =  D_T \nabla f_T + \mathcal{Y}\nabla f_L + j_s f_L,\label{eq: j_T}
\end{align}
where $f_{L(T)}(\epsilon)$ is symmetric (antisymmetric) in energy part of the electron distribution function, and the kinetic coefficients in the $\theta$-representation read
\begin{align}
&D_{L/T} = \frac{1}{2} (1 +  | \cosh \theta |^2 \mp |\sinh \theta|^2 \cosh ( 2\Imag{\chi} ))\\
&\ \mathcal{Y} = \frac{1}{2}|\sinh \theta|^2\sinh ( 2\Imag{\chi} ),\  j_s = \Imag{j_E}.
\end{align}
Note that the spectral function ${\cal Y}$ accounts for electron-hole asymmetry in our system. At this point it is worth mentioning here that generic 
metallic structures without superconductors normally exhibit very weak thermoelectric effect due to the presence of electron-hole symmetry in such 
structures.

Eqs.~(\ref{eq: Theta})--(\ref{eq: j_T}) need to be supplemented with proper boundary conditions at all interfaces of our structure. 
Here we will restrict our analysis to the important limit of fully transparent metallic interfaces meaning that at all wire nodes (i) the functions 
$\theta,\ \chi,\ f_L$ and $f_T$ remain continuous and (ii) the spectral currents ${\cal A}\nabla \theta  ,\ {\cal A}  \nabla \chi, \ {\cal A} j_L$ and 
${\cal A} j_T$ are conserved. At the interfaces between the wires and the N-terminals the functions $\theta$, $\chi$, $f_L$ and $f_T$ take
the bulk values deep inside the terminals. At NS boundaries for $| \epsilon | < \Delta$ we have $j_L = 0$ and $f_T = 0$ just implying that (a) subgap quasiparticles transfer no heat across the interface and (b) charge imbalance vanishes at the NS boundary.

\section{Thermopower}
\label{sec: Thermo}
The task at hand is to study the thermoelectic effect between the reservoirs N$_3$ and N$_4$.
Let us set $l_{N,3} = l_{N,4} = l_N$. In this case by symmetry we have $V_3 = V_4 = V_N$ provided all terminals are kept at the same temperature $T$. The potential $V_N$ just represents the voltage drop at the point where the terminals N$_3$ and N$_4$ are attached to the central wire $l_c$.

In what follows we are going to evaluate the thermoelectric potential difference $V_3-V_4$. Let us define the thermopower as
\begin{equation}
{\cal S}_{34} = \frac{V_3-V_4}{T_3-T_4}\Big|_{T_4 \mapsto T_3}.
\label{defS}
\end{equation}
We will demonstrate that this quantity obeys the following general formula
\begin{equation}
{\cal S}_{34} (V, \phi) = {\cal S}_0(V) + {\cal S}_{\rm odd}(V,\phi) + {\cal S}_{\rm even}(V,\phi),
\label{S34}
\end{equation}
where ${\cal S}_0(V) = \langle{\cal S}_{34}\rangle_\phi$ and the last two terms represent respectively odd and even in $\phi$ oscillating $2\pi$-periodic functions  with zero average over $\phi$. Hence, in the agreement with our previous work \cite{PD18} the thermopower ${\cal S}_{34} (V, \phi)$ (\ref{S34}) is, in general, neither odd nor even function of $\phi$ taking a nonzero value at $\phi =0$.

\begin{figure}[!ht]
\includegraphics[width=\columnwidth]{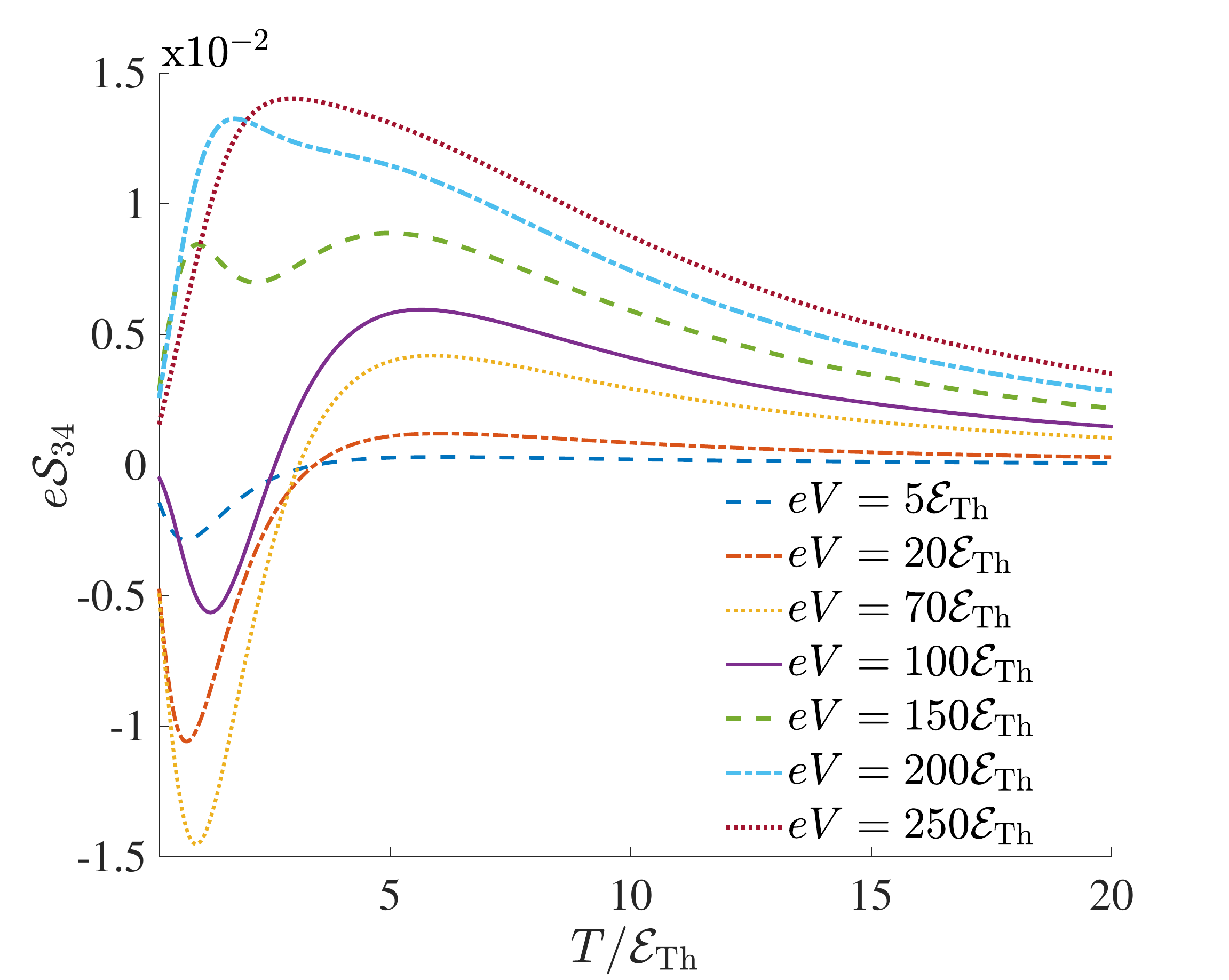}
\caption{The thermopower $e\mathcal{S}_{34}$ for $\phi = 0$ as a function of temperature at different voltage bias values. Here we set $l_{S,1} = l_{S,2} = l_{N,1} = l_{N,2}=  1/3 L,\ l_c = 0.1 L + (1/3 - 0.1) L,\ l_{N,3} = l_{N,4} = 1/2 L$ and $\Delta = 10^3\mathcal{E}_{\rm Th}$.}
\label{fig: TempDep_phi_0}
\end{figure}

Making use of the formalism outlined in the previous section we evaluated the thermopower ${\cal S}_{34} (V, 0)$
numerically as a function of temperature. The results of this calculation are displayed in Fig.~\ref{fig: TempDep_phi_0}. We observe a rather non-trivial non-monotonous behavior of the thermopower as a function of $T$ which -- depending on the bias voltage -- can even change its sign. At high temperatures strongly exceeding the Thouless energy $\mathcal{E}_{\rm Th}$ the thermopower ${\cal S}_{34}$ shows a slow (power-law) decay with increasing $T$ which is reminiscent of that for the even in $\phi$ (Aharonov-Bohm) current component~\cite{GWZ97}.

Turning to the phase dependence of the thermoelectric signal, one can resolve the kinetic equations \eqref{eq: j_L} and \eqref{eq: j_T}. Then inside the normal wires $l_{N,3}$ and $l_{N,4}$ one finds
\begin{eqnarray}
&&\begin{pmatrix}
j_L^{N_{3,4}} \\ j_T^{N_{3,4}}
\end{pmatrix}
=
\dfrac{\hat{\bm{G}}^{N_{3,4}}}{\mathcal{A}\sigma_N}
\begin{pmatrix}
f_L^N(V_{3,4}) - f_L^c \\ f_T^N(V_{3,4}) - f_T^c
\end{pmatrix},\label{eqn: G_matrix}\\
&&f_{L/T}^N(V) = \frac{1}{2}\Big( \tanh\frac{\epsilon + eV}{2T} \pm \tanh\frac{\epsilon - eV}{2T} \Big).
\end{eqnarray}
Here $f_{T/L}^c$ are the components of the distribution function evaluated in the central crossing point, $\sigma_N$ is the normal conductivity of the wire and $\hat{\bm{G}}^{N} = \hat{\bm{G}}^{N_{3}}=\hat{\bm{G}}^{N_{4}}$ is the spectral conductance matrix defined as
\begin{equation}
\hat{\bm{G}}^{N}=
\begin{pmatrix}
\bm{G}^{N}_L & - \bm{G}^{N}_{\mathcal{Y}} \\
\bm{G}^{N}_{\mathcal{Y}} & \bm{G}^{N}_T
\end{pmatrix}
=
\dfrac{\mathcal{A}\sigma_N}{\displaystyle
\int_{l_{N}}
\begin{pmatrix}
D_L & -\mathcal{Y} \\
\mathcal{Y} & D_T
\end{pmatrix}^{-1}
dx
}.
\label{condG}
\end{equation}
Here the integration is performed either over the wire $l_{N,3}$ or $l_{N,4}$. In the absence of superconductivity this matrix reduces to the diagonal one proportional to the conductance of the corresponding normal wire. Assuming the proximity effect deep inside the wires $l_{N,3}$ and $l_{N,4}$ to be sufficiently weak one can rewrite Eq.~\eqref{condG} in the form
\begin{equation}
\hat{\bm{G}}^{N}
\approx
\dfrac{\mathcal{A}\sigma_N}{l^2_{N}}
\displaystyle
\int_{l_{N}}
\begin{pmatrix}
D_L & -\mathcal{Y} \\
\mathcal{Y} & D_T
\end{pmatrix}
dx.
\label{condG2}
\end{equation}
Bearing in mind that no current can flow into and out of electrically isolated terminals N$_{3}$ and N$_{4}$ and by using the $T$-component of 
Eq.~(\ref{eqn: G_matrix}), we transform the condition $\displaystyle\int d\epsilon \{j_T^{N_{3}} - j_T^{N_{4}}\} = 0$ to:
\begin{multline}
\int d \varepsilon
\Bigl\{
\bm{G}^{N}_T [f_T^N(V_3) -f_T^N(V_4)]
+\\+
\bm{G}^{N}_\mathcal{Y} [f_L^N(V_3) -f_L^N(V_4)]
\Bigr\}
=0.
\end{multline}
This equation determines the relationship between the terminal temperatures $T_3,\ T_4$ and the induced thermoelectric voltages $V_3,\ V_4$. Assuming $\delta T = T_3 - T_4$ to be sufficiently small and evaluating the  voltage difference $V_3-V_4$ up to linear in $\delta T$ terms, from Eq. (\ref{defS})
we obtain
\begin{equation}
e\mathcal{S}_{34} \approx
\dfrac{1}{4T_N^2}
\dfrac{l_{N}}{\mathcal{A}\sigma_N}
\int
d \varepsilon
\dfrac{
(\bm{G}_T + \bm{G}_\mathcal{Y}) (\varepsilon + e V_N)
}{\cosh^2 [(\varepsilon + e V_N)/(2T_N)]}.
\label{s34}
\end{equation}
Here $V_N$ is the induced voltage in both terminals N$_3$ and N$_4$ for $T_3 = T_4 = T_N$, i.e. in the absence of the temperature gradient. It is also important to keep in mind that the spectral conductance $\bm{G}_T (\varepsilon, \phi)$ ($\bm{G}_\mathcal{Y} (\varepsilon, \phi)$) in Eq. (\ref{s34}) is an even (odd) function of both $\varepsilon$ and $\phi$.

Eq.~\eqref{s34} represents the cental result of this work.

The first observation to be made from this result is that the thermoelectric effect vanishes in the symmetric geometry with $l_{S,1} = l_{S,2},\ l_{N,1} = l_{N,2}$ and with the terminals N$_3$ and N$_4$ attached to the center. It is easy to see that in this case we have $V_N = 0$ and, hence, the even term in Eq.~\eqref{s34} vanishes after the energy integration. At the same time, the spectral function ${\cal Y}$ remains zero in the wires $l_{N,3,4}$, thus providing $\bm{G}_\mathcal{Y} = 0$.
\begin{figure}
\centering
\includegraphics[width=1\linewidth]{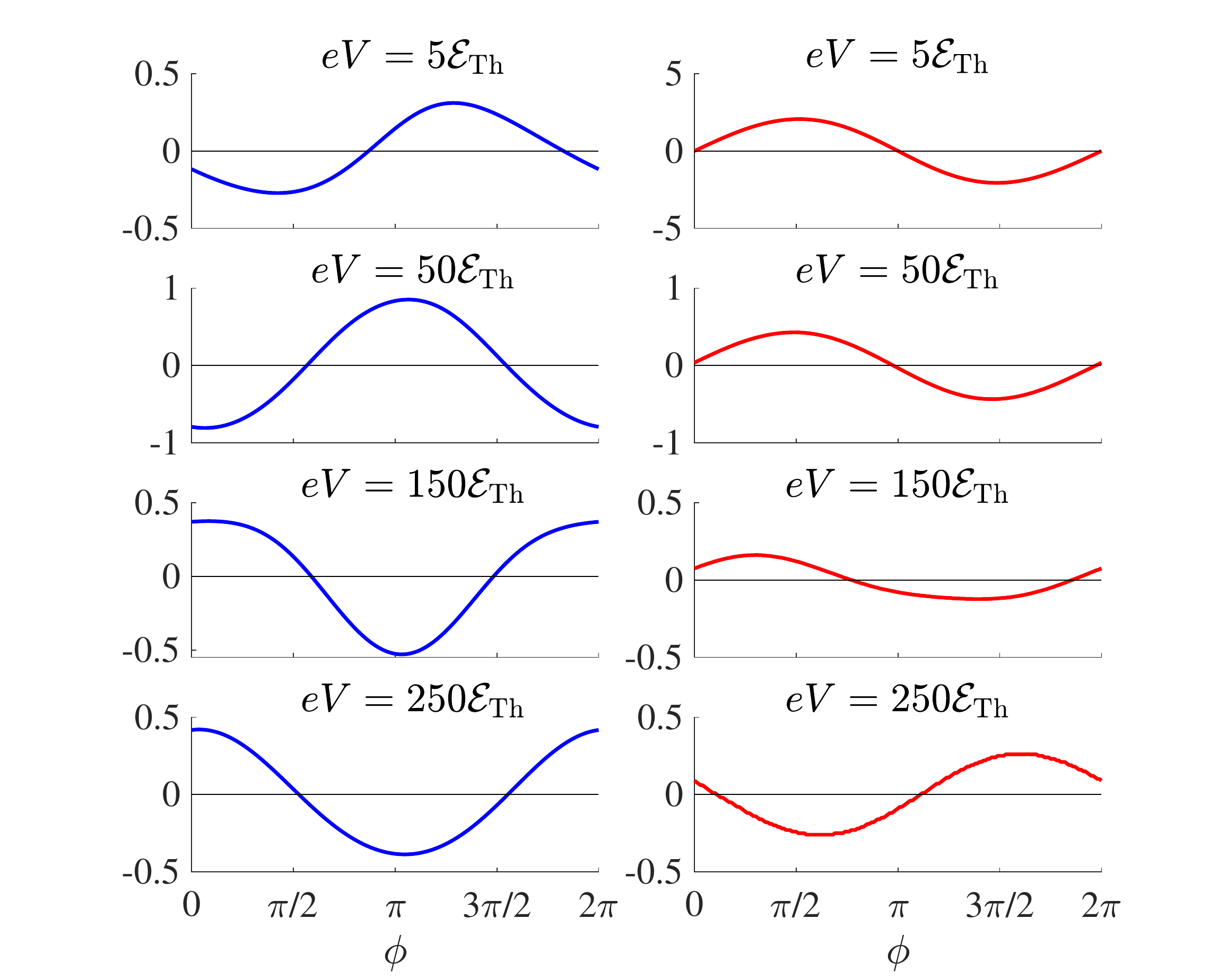}
\caption{The oscillating parts of the thermopower (left panel) and the current $I_S$ (right panel) as functions of $\phi$ at different bias voltages $V$. The thermopower is measured in units of $10^{-2}/e$, while we express the current as $eR_L I_S/{\cal E}_{\rm Th}$ with $R_L =  L/(\sigma_N{\cal A})$.}
\label{fig: MainRes}
\end{figure}

In addition, Eq.~\eqref{s34} implies that at low voltages, such that $eV_N \lesssim \mathcal{E}_\mathrm{Th}$, the odd in $\phi$ contribution is dominant to the thermopower. (Note that $eV_N \lesssim \mathcal{E}_\mathrm{Th}$ does not necessarily mean $eV \lesssim \mathcal{E}_\mathrm{Th}$.) At larger voltages $e V_N \gtrsim \mathcal{E}_\mathrm{Th}$, the even harmonics starts to dominate, cf. Fig.~\ref{fig: MainRes}. The latter observation is specific to our geometry, where at such voltages the odd harmonics becomes suppressed by the factor $\sim (l_{c,1} - l_{c,2})/L$.

In Fig.~\ref{fig: MainRes} (left panel) we display the thermopower-phase relations at $T = \mathcal{E}_{\rm Th}$ and for several voltage-bias values $V$. We observe that, while at lower voltages the thermopower $\mathcal{S}_{34} (\phi)$ is (almost) odd in $\phi$, it then tends to an even function of $\phi$ rather quickly with increasing $V$. It is interesting to compare this behavior of the thermopower with that of the phase-dependent current $I_S(\phi)$ flowing between the two superconducting terminals. The current $I_S(\phi)$
was recently evaluated elsewhere \cite{PD18J}, and the corresponding results are displayed in the right panel of
Fig.~\ref{fig: MainRes}.

\begin{figure}[!ht]
\includegraphics[width=\columnwidth]{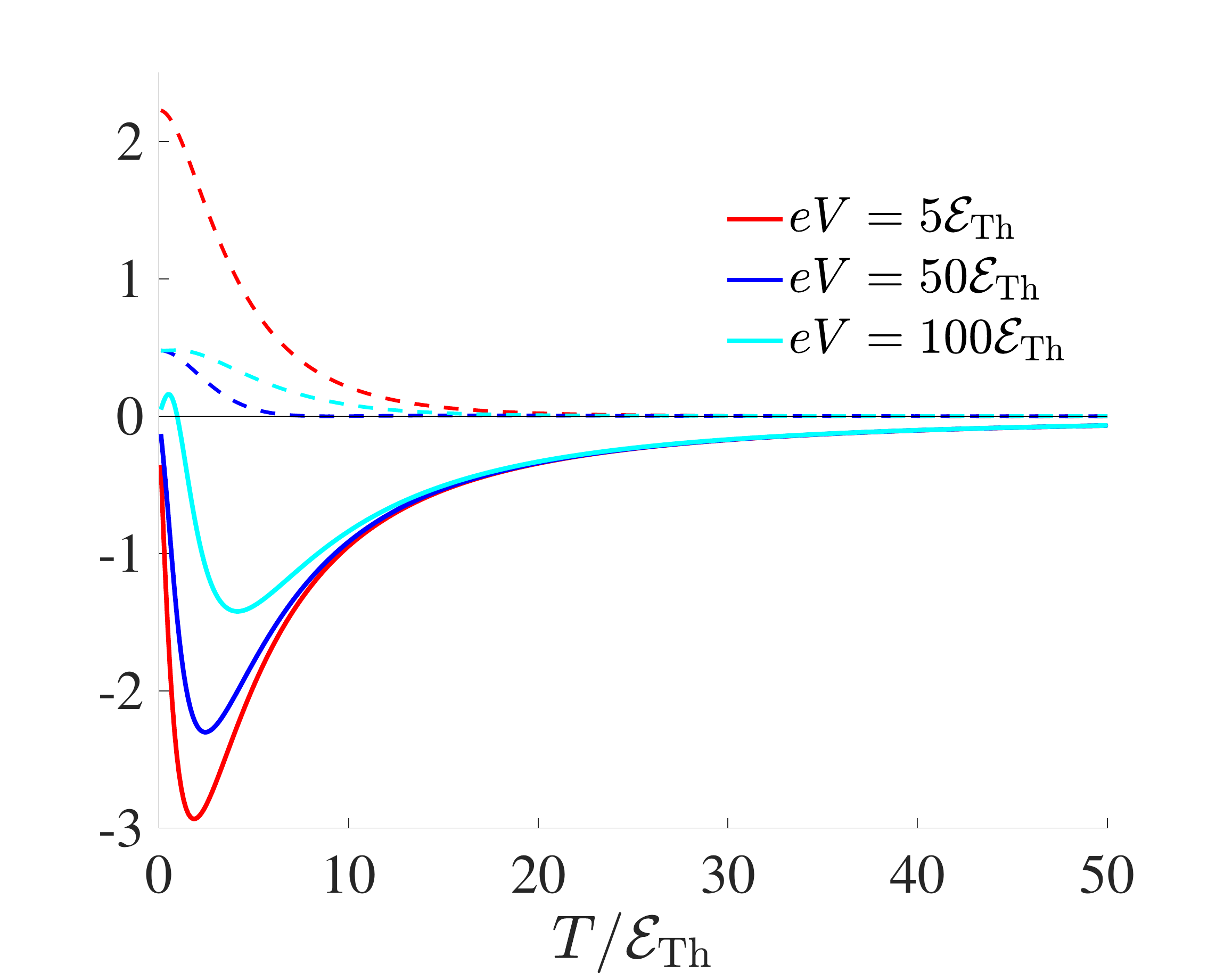}
\caption{The principal odd Fourier harmonics of the thermopower-phase (solid lines) and the current-phase (dashed lines) relations as functions of temperature for different bias voltages. The thermopower values are normalized to $10^{-3}/e$, while the current $I_S$ is expressed in the same manner as in Fig.~\ref{fig: MainRes}. With increasing $T$ the current amplitude decays much faster as compared to that of ${\cal S}_{34}$.}
\label{fig: Q_Thermo_IS}
\end{figure}

In contrast to the thermopower $\mathcal{S}_{34} (\phi)$, the odd in $\phi$ harmonics dominates the function $I_S(\phi)$ at all given values of the bias voltage $V$. This observation indicates that the current $I_S(\phi)$ and the thermopower $\mathcal{S}_{34} (\phi)$ may have a {\it different origin}. Indeed, the current $I_S(\phi)$ is defined through the spectral supercurrent $j_E$, while the odd harmonics of $\mathcal{S}_{34} (\phi)$ originates from the spectral function $\mathcal{Y}$. To further stress the difference between these two quantities, in Fig.~\ref{fig: Q_Thermo_IS} we plot the temperature dependence of the odd harmonics for both the current and the thermopower.  We observe that at sufficiently high temperatures $T\simeq 15{\cal E}_{\rm Th}$ the current amplitude is already completely suppressed, while the thermopower amplitude is not.

Note that in our previous study~\cite{PD18} we addressed the setup with a somewhat different topology from that analyzed here. In ~\cite{PD18} we found that the thermopower-phase relation closely follows the dependence $I_S(\phi)$ at all values of $V$. Furthermore, for $V=0$ it was argued~\cite{VH} that the thermopower ${\cal S}$ is simply proportional to ${\cal S} \propto d I_S/dT$. Making use of this relation one could be tempted to speculate that both the supercurrent and the thermopower (i) have the same origin and (ii) are odd functions of $\phi$. The results displayed in Fig.~\ref{fig: Q_Thermo_IS} clearly demonstrate that the relation ${\cal S}_{34} \propto d I_S/dT$ {\it does not hold} for Andreev interferometers under consideration.

Thus, the above analysis allows us to conclude that the system topology -- along with such parameters as $V$, $T$ and $\mathcal{E}_{\rm Th}$ -- plays a decisive role for the thermopower-phase relation. This conclusion qualitatively agrees with experimental observations \cite{Venkat1}. Our results demonstrate that -- depending on the system topology -- the behavior of the thermopower can be diverse: In symmetric setups it may get small or even vanish, while in non-symmetric ones the origin of the odd-part of the thermopower-phase relation may be determined either by the Josephson-like effect or through the spectral function ${\cal Y}$ which accounts for electron-hole asymmetry in our system.

This work was supported in part by RFBR Grant No. 18-02-00586. P.E.D. acknowledges support by Skoltech NGP Program (Skoltech-MIT joint project).


\begin{thebibliography}{99}
\bibitem{C1} Throughout this paper we set both Boltzmann and Planck constants equal to unity $k_B = \hbar = 1$.
\end{thebibliography}
\end{document}